\documentclass[runningheads]{llncs}
\usepackage[ruled]{algorithm2e}
\usepackage{xcolor}
\usepackage{amsmath}
\usepackage{amssymb}
\usepackage{float}
\usepackage{graphicx}
\usepackage[square,sort,comma,numbers]{natbib}
\usepackage{hyperref}
\usepackage{xparse}

\SetKw{Continue}{continue}
\SetKw{Throw}{throw}
\SetKwFor{uWhile}{while}{do}{}
\SetKwFor{endWhile}{}{}{end}

\definecolor{CadetGrey}{rgb}{0.57, 0.64, 0.69}
\definecolor{ColourBlindCyan}{RGB}{51,187,238}
\definecolor{ColourBlindOrange}{RGB}{238,119,51}
\definecolor{ColourBlindTeal}{RGB}{0,153,136}
\newcommand{\pdfspace}[1]{{\color{ColourBlindOrange}#1}}
\newcommand{\pdftoken}[1]{{\color{ColourBlindTeal}#1}}

\newcommand{\bit}{\,bit}
\newcommand{\byte}{\,B}
\newcommand{\pt}{\,pt}

\newcommand{\aenc}{\textsc{ae}}
\newcommand{\bch}{\textsc{bch}}

\newcommand{\docid}{\texttt{document\_id}}
\newcommand{\downid}{\texttt{download\_id}}
\newcommand{\ecc}{\textsc{ecc}}
\newcommand{\Enc}{\texttt{Enc}}
\newcommand{\ewiser}{\textsc{ewiser}}
\newcommand{\ic}{\textsc{ic}}
\newcommand{\id}{\textsc{id}}
\newcommand{\indcpa}{\textsc{ind-cpa}}

\newcommand{\iso}{\textsc{iso}}
\newcommand{\lcs}{\textsc{lcs}}
\newcommand{\mac}{\textsc{mac}}
\newcommand{\nlp}{\textsc{nlp}}
\newcommand{\nltk}{\textsc{nltk}}
\newcommand{\ocr}{\textsc{ocr}}
\newcommand{\pdf}{\textsc{pdf}}
\newcommand{\Pdf}{\textsc{Pdf}}
\newcommand{\pos}{\textsc{pos}}
\newcommand{\swico}{\textsc{swico}}
\newcommand{\Tag}{\texttt{Tag}}
\newcommand{\twico}{\textsc{twico}}
\newcommand{\utc}{\textsc{utc}}

\def\ifempty#1{\def\temp{#1}\ifx\temp\empty}

\DeclareMathOperator*{\argmax}{arg\,max}

\newcommand{\Call}[2]{\ifempty{#2}\texttt{#1}\else\func{\texttt{#1}}{#2}\fi}

\makeatletter
\newcommand{\func}{\@ifstar{\func@star}{\func@nostar}}
\newcommand{\func@nostar}[2]{\ensuremath{\operatorname{#1}\mathopen{}\left(#2\right)\mathclose{}}}
\newcommand{\func@star}[2]{\ensuremath{#1\mathopen{}\left(#2\right)\mathclose{}}}
\makeatother
\newcommand{\funcname}[1]{\operatorname{#1}}

\newcommand{\idx}[1]{\mathopen{}\left[ {#1} \right]\mathclose{}}

\newcommand{\isotime}[2]{{#1}T{#2}Z}

\NewDocumentCommand{\pe}{m O{1}}{\ensuremath{#1 \gets #1 + \ifempty{#2}1\else#2\fi}}

\begin{document}

\title{Traitor-Proof PDF Watermarking}

\author{Fabrizio d'Amore\inst{1}\orcidID{0000-0002-6518-2445} \and
Alessandro Serpi\inst{2}}

\authorrunning{F. d'Amore and A. Serpi}

\institute{Sapienza Univ.\ di Roma, \\
Dip.\ Ingegneria informatica, automatica e gestionale A. Ruberti, 
Italia\\ 
\url{https://www.diag.uniroma1.it/en}\\
\email{damored@diag.uniroma1.it}
\and\
Independent researcher, \\
Roma, Italia\\
\email{serpi.1647244@studenti.uniroma1.it}
}

\maketitle

\begin{abstract}
  This paper presents a traitor-tracing technique based on the watermarking of digital documents (\pdf{} files in particular).
  The watermarking algorithm uses a chain of three separate techniques that work in synergy.
  The embedded payload can withstand a wide range of attacks and cannot be removed without invalidating the credibility of the document.
  
  We will present an implementation of the approach and discuss its limitations with respect to documents that can be watermarked and quality of the watermarked documents.
  We will also analyse two payload alternatives and how the encryption scheme may alleviate the chilling effect on whistle-blowing.
\keywords{Text watermarking, Traitor tracing, Pdf}
\end{abstract}

\section{Introduction}
All organisations store and process sensitive documents whose unauthorised disclosure would result in a financial and credibility loss for the organisation itself.
Given their importance, usually those documents are stored as \pdf{} files and are accessible only by specific vetted users.
Therefore, they are quite safe from both external and internal attacks.
However, users that are legitimately authorised to access sensitive documents may want to leak them; take for example disgruntled employees in search of revenge or whistle-blowers that failed to start a legal investigation.
Thus, organisations need a \pdf{} watermarking algorithm that invisibly embeds a robust payload into sensitive documents to identify the user that downloaded and divulged them.
In other words, organisations need a traitor-tracing scheme based on digital watermarks.

There are several approaches to invisible document watermarking, each with its own advantages and vulnerabilities.
In this paper, we will present a novel traitor-tracing scheme that implements three different approaches to withstand to a wide range of attacks.

We will explain the workflow required to produce a document visually similar to the original and what limitations it causes.
We will also analyse two payload alternatives and how to alleviate the chilling effect on whistle-blowing.

\subsection{Related Works}
There are several approach to document watermarking.
The simplest technique that modifies the document itself is to embed information into the file's metadata.
However, it is not robust as the payload is easily erased.
We can consider the document as a series of images that can be watermarked with general-purpose techniques or specific for textual documents.
The watermark withstands printing, but it is eliminated if the adversary employs \ocr{} or re-types the document.

Font-based algorithms deserve a separate analysis thanks to their heterogeneity.
They can be considered as image-based techniques with special properties.
\cite{US20040001606A1} was the first to use altered character glyphs to watermark textual documents.
They proposed two techniques, one changes the glyphs in specific locations, the other changes all glyphs of specific characters \cite{US20040001606A1}.
Only the latter withstands partial document erasure, but its capacity is lower than the former's.
\cite{10.1145/1161366.1161387} were the first to treat fonts as peculiar drawings and to
design a watermarking algorithm that took advantage of their intrinsic characteristics by changing the vector curves \cite{10.1145/1161366.1161387}.
Equivalent techniques have been proposed for non-Latin characters such as Arabic \cite{10.5121/ijaia.2015.6401} and Chinese \cite{tan-et-al}.

A structural watermarking algorithm alters the formatting in order to embed the payload.
\cite{10.1109/49.464718} were the first to describe algorithms that modify inter-word and inter-line spaces \cite{10.1109/49.464718, brassil-et-al2}.
\cite{10.1109/76.974678} proposed an algorithm that uses average inter-word spaces in lines \cite{10.1109/76.974678}.
\cite{10.1109/ICDAR.2003.1227767} generalised the concept to arbitrary inter-word space statistics and increased robustness by using word classes \cite{10.1109/ICDAR.2003.1227767}.
Structural watermarks have the same advantages and drawbacks as image-based ones\footnote{In fact, they can be considered as highly specialised image-based watermarking algorithms.}.
The payload withstands printing, but not \ocr{} and re-typing.
In addition, if the adversary has the document in the original format, he/she can easily eliminate the payload.

Linguistic watermarking algorithms act directly on the text.
\cite{10.1007/3-540-36415-3_13} were the first to propose semantic and syntactic watermarking schemes \cite{10.1007/3-540-36415-3_13}. \cite{10.1145/1161366.1161397} designed syntactic algorithms that encode a payload using substitution of synonyms \cite{10.1145/1161366.1161397} and abbreviations, acronyms, and typographical errors \cite{10.1117/12.706980}. \cite{10.1117/12.708111} analysed several tools for syntactic watermarking \cite{10.1117/12.708111} and proposed morpho-syntactic editing \cite{10.1016/j.csl.2008.04.001}.

\section{Preliminaries}
Document watermarking techniques can be classified into blind, semi-blind and non-blind schemes based on the information required to extract the payload $p$ from a watermarked document $d_w$.
Blind algorithms need only the watermarked document (and the secret key $k$, if present).
Non-blind algorithms require in addition the original document $d$. 
Semi-blind algorithms are slightly different because they check whether a specific payload was embedded into a watermarked document.

Invisible watermarking assumes an active adversary whose objective is to eliminate the watermark without compromising the document's integrity.
The actions that the adversary can perform are usually divided into content- and formatting-based attacks.
In the former category fall all actions that alter the information contained in the document, such as synonym substitution, syntactic transformations, and word insertion, deletion and rearrangement.
In the latter category fall all actions the alter the presentation of the document, such as font modification, printing, \ocr{}, retyping, and rotation.

Since Linguistic watermarking algorithms modify the text, they withstand all formatting-based attacks.
However, depending on the algorithm, the payload may be lost if the adversary performs a content-based attack.
Structural algorithms are usually more fragile.
Nonetheless, they may be preferable in some contexts.
Since they do not alter the text, therefore the meaning of the document is surely preserved.

Linguistic algorithms require a lexical database to represent the language of the document.
WordNet \cite{10.1145/219717.219748} is among the most used.
Originally developed in English in the Cognitive Science Laboratory of Princeton University, it has now expanded to more than 200 languages.
WordNet links words (sequence of characters) and their senses into semantic relations including synonyms, hyponyms, and meronyms.
Synonyms are grouped into `synsets' (or synonym rings) with short definitions and usage examples.

Portable Document Format (\pdf) is the \iso{} standard for the representation of electronic documents \cite{ISO32000-2:2017}. Documents are presented always in the same manner, independently from the user's system.
A \pdf{} file includes at least four elements: the header, the body, the cross-reference table and the trailer.
In particular, the body of a document represents the file’s contents and is constituted by a sequence of indirect objects (i.e. objects that are stored elsewhere in the file with respects to where they are referenced).

\Pdf{} defines several operators to dictate how text is showed and positioned in a document.
The \texttt{Tf} operator specifies the font.
\texttt{Tc} and \texttt{Tw} define respectively character and word spacing.
\texttt{Tj} shows a text string based on the aforementioned parameters.
\texttt{TJ} is the more advanced version of \texttt{Tj}, as it can control the positioning of individual characters inside a string.

\subsection{Word Similarity}\label{sec:word-sim}
There are several similarity functions that analyse the closeness of two senses, many of which are based on WordNet.
Let $\Call{lcs}{x, y}$ be a function returning the least common subsumer (i.e.\ the most specific ancestor) of senses $x$ and $y$, let $\Call{ic}{x}$ be a function returning the information content (\ic) of sense $x$, and let $\Call{depth}{x}$ be the function returning the depth of node $x$. 
Obviously, each of these functions returns the same value for all senses in a synset.

The Wu-Palmer similarity considers the depth of the  \lcs{} in relations with the depth of the sense. It goes from 0 to 1 and is defined as \[
	\Call{wup}{\langle w_x, s_x \rangle, \langle w_y, s_y \rangle} \triangleq \frac{2 \cdot \Call{depth}{\Call{lcs}{\langle w_x, s_x \rangle, \langle w_y, s_y \rangle}}}{\Call{depth}{\langle w_x, s_x \rangle} + \Call{depth}{\langle w_y, s_y \rangle}} \, .
\]
The Leacock-Chodorow scales the depth of the \lcs{} with respect to the maximum depth of the taxonomy in which the senses occur: \[
	\Call{lch}{\langle w_x, s_x \rangle, s_y} \triangleq \log \frac{\Call{len}{\Call{path}{\langle w_x, s_x \rangle, \langle w_y, s_y \rangle}}}{2 \cdot \Call{maxLength}{\langle w_x, s_x \rangle, \langle w_y, s_y \rangle}} \, .
\]
The Resnik similarity corresponds to the information content of the least common subsumer: \[
	\Call{res}{\langle w_x, s_x \rangle, \langle w_y, s_y \rangle} \triangleq \Call{ic}{\Call{\lcs}{\langle w_x, s_x \rangle, \langle w_y, s_y \rangle }} \, .
\]
Its lower bound is 0 whereas the upper bound is $\log N$, where $N$ is the number of senses in the taxonomy.
The Jiang-Conrath similarity is defined as \begin{align*}
	\Call{jcn}{\langle w_x, s_x \rangle, \langle w_y, s_y \rangle} \triangleq & \left[ \Call{ic}{\langle w_x, s_x \rangle} + \Call{ic}{\langle w_y, s_y \rangle} \right. \\ & \phantom{\left[\right.} \left. \mathrel{-} 2 \cdot \Call{ic}{\Call{\lcs}{\langle w_x, s_x \rangle, \langle w_y, s_y \rangle}} \right]^{-1} .
\end{align*}
If the senses are in the same synset, $\funcname{jcn}$ encounters a singularity.
However, $+\infty$ is not returned even in those environments that supports it, such as Python and R.
Instead, 1 is returned.
Lin similarity is another extension of Resnik's: \[
	\Call{lin}{\langle w_x, s_x \rangle, \langle w_y, s_y \rangle} \triangleq \frac{2 \cdot \Call{ic}{\Call{\lcs}{\langle w_x, s_x \rangle, \langle w_y, s_y \rangle}}}{\Call{ic}{\langle w_x, s_x \rangle} + \Call{ic}{\langle w_y, s_y \rangle}} \, .
\]
By convention, all three information-content-based similarity functions span from 0 to 1.

Path-based similarities (Wu-Palmer and Leacock-Chodorow) cannot be meaningfully compared  to \ic-based similarities.
Nonetheless, based on empirical measures, we know that Leacock-Chodorow is more respondent to reality than Wu-Palmer and Resnik performs worse than Jiang-Conrath and Lin \cite{10.1109/ICCASM.2010.5619038}.

\section{Approach}
Traitor tracing has specific assumptions and requirements.
The original work cannot be destroyed because a new watermarked version needs to be generated at each access.
In addition, we assume to be working with official documents that need to be preserved.
They can be discarded only when they cease to be relevant, which is when the traitor tracing mechanism is not required any more.
Therefore, we can safely use both blind and non-blind watermarking schemes.
Semi-blind algorithms are not suited to traitor tracing because the embedded payload is not known when checking a leaked document.

Another vital requirement is high embedding effectiveness.
The used algorithm should reliably embed all possible payloads in all admissible objects.
Obviously, this result is unattainable.
In real life, textual works may be even a few words long, thus impossible to be dependably watermarked by a robust technique.
Since we use three watermarking algorithms, it is probable that at least one of them (especially the font-based one) will successfully embed the payload.

This consideration takes us to the next requirement: robustness.
We do not distinguish between normal content processing and deliberate attacks because the result is the same: watermark removal.
Since we conform to Shannon's maxim, we assume that the adversary has complete knowledge of the used algorithm except for the secret key.
Obviously, when one or more algorithms fail to embed the payload, robustness decreases because they cannot withstand some types of attack that it normally could.

The document watermarking technique we propose is composed by three separate algorithms that work in synergy: the first is linguistic, the second is structural and the third is font-based.
We designed and implemented this technique based on the \pdf{} 2.0 specification.
Nonetheless, it can be expanded to other formats with new document parser/embedder algorithms.

\subsection{Adversary}
The proposed algorithm was designed under the assumption that the adversary must leak a credible file to maintain their reputation.
As such, they can either publish the entire processed document (after having performed formatting- and content-based attacks) or just a portion visually identical to the original.
This behaviour is inspired by the code of conduct of real-world leakers such as WikiLeaks or the press.
However, it fails to capture more prosaic cases, such as disgruntled employees.
Those adversaries do not care about neither credibility nor reputation.
Their only goal is to cause the maximum damage to their (possibly former) employer.
Since they increase the set of allowed attacks almost limitlessly, they were excluded from the chosen adversary model.

\subsection{Linguistic}
A homograph is a word that is spelled exactly as another word but has different meaning\footnote{Some dictionaries impose that the two words must also have different origin. However, it is not relevant in this context.}.
For example, `bear' can signify both the carnivore mammal and to carry.
Thus, a homograph belongs to multiple synsets.
The use of homographs for text watermarking was first proposed by \cite{10.1145/1161366.1161397} \cite{10.1145/1161366.1161397}.
Performing synonym substitution with homographs increases ambiguity.
Consequently, erasing the payload by restoring the original words is more difficult.
While the algorithm proposed work is quite robust, it was not specifically designed for traitor tracing.
As such, we can design a new algorithm that is better suited to our use case.

While we use WordNet to define the watermarking technique, it is not a fixed requirement: we can seamlessly replace it with any other framework that uses a similar concept.
Better yet, organisations can use natural language processing to create synsets from a specific corpus in order to adapt the algorithm to their specialised jargon.

\subsubsection{Graph Creation}
Let $G$ be a graph where vertices represent $\langle$word, sense$\rangle$ pairs and edges indicate that their endpoints are in the same synset.
Using $G$ and a secret key $k_G$, we create a new weighted graph $G_w$.
This step is computationally expensive, but is performed a single time for all embedding and extractions that use the same WordNet version and $k_G$.
Moreover, $G$ and edge weights are independent of the secret key.
Therefore, they can be re-used in all computations of $G_w$ that are based on the same $G$.

We collapse all vertices with the same word representation in a single vertex.
We leave the edges as they are: two vertices are linked if and only if their senses share at least a synset.
Then, we assign a weight to the vertices that reflects the similarity between the endpoints.
Let $\func*{S}{x}$ be a function that returns all senses of word $x$.
Given two neighbours $x$ and $y$, the weight of the edge linking them is
\[
	\Call{weight}{x, y} \triangleq \frac{ \sum_{s_x \in \func*{S}{x}} \sum_{s_y \in \func*{S}{y}} \Call{sim}{\langle x, s_x\rangle, \langle y, s_y \rangle}}{|\func*{S}{x}| \cdot |\func*{S}{y}|} \, ,
\] 
where $\funcname{sim}$ is a similarity function between two senses.
There are several such metrics, see sec. \ref{sec:word-sim} for a brief review of some of them.
Functions whose codomain is $[0, 1]$ are the most appropriate for the watermarking algorithm.
Therefore, the two choices among those analysed are Jiang-Conrath and Lin.
While in our implementation we chose the latter, the former is still acceptable.

Finally, we label all homographs with 0 or 1 in a way that each word has approximately the same number of 0- and 1-homographs as neighbours.
For example, we could use the least significant bit of a cryptographic hash function with key $k_G$.
If the hash function is resistant to preimage attacks, then the watermarking algorithm withstands to Single-Watermarked-Image-Counterfeit-Original (\swico) and Twin-Watermarked-Images-Counterfeit-Original (\twico) attacks.
Even though such a property is not necessary in our use case, it may still be desirable.
In order to complicate the adversary's task, we can label the edges instead of the vertices.
In this way, the hash function can accept the concatenation of the endpoints as input.

Since $G_w$ is created deterministically, it is not a problem if an attacker compromises the framework and erases $G_w$, as long as $G$ and $k_G$ are safely stored or can be re-computed.

\subsubsection{Embedding}
Let $\Call{neighbours}{x, b}$ be the function that returns all neighbours of $x$ labelled with $b$.
Then, alg. \ref{alg:linguistic-embed} embeds a payload $p'$ in a document $d$.
To streamline the pseudocode, the algorithm is in-place.
Since the \pdf{} parser and embedder is a distinct algorithm, this simplification has no practical consequences.

\begin{algorithm}[h]
    \caption{Linguistic embedding.}
	\label{alg:linguistic-embed}
	
	\DontPrintSemicolon{}
	\KwData{original document $d$ as an array of words;\newline
	    payload $p'$ as defined in sec. \ref{sec:encryption};\newline
	    array $u$ containing whether the corresponding word in $d$ can be replaced;
	}
	\KwResult{whether the embedding succeeded}
	
	\BlankLine{}
	$i \gets 0$ \tcp*{counter for $p'$}
	$j \gets 0$ \tcp*{counter for $d$}
		
	\BlankLine{}
	\While{$i < |p'|$}{
		\If{$j = |d|$}{
		\tcp{document ended but}
		\tcp{$p'$ was not fully embedded}
		\KwRet{$\bot$}}\;
	
		\If{$d\idx{j} \notin G_w \vee u\idx{j}$}{
		    \tcp{word cannot be replaced}
			$j \gets j + 1$\;
			\Continue\;
		}
		
		\BlankLine{}
		$\displaystyle d\idx{j} \gets \argmax_{w \in \func*{neighbours}{d\idx{j},\, p\idx{i}}} \Call{weight}{d\idx{j},\, w}$\;
		
	    \pe{i}\;
		\pe{j}\;
	}
	
	\BlankLine{}
	\tcp{embed the error-correcting code}
	\tcp{see sec. \ref{sec:ecc})}
	\Call{embedEcc}{d,\, p',\, j + 1}\;
	\KwRet $\top$\;
\end{algorithm}

Even though the algorithm as presented is perfectly suitable for documents with no strict formatting, it may not produce acceptable results when applied to justified documents.
In fact, we are not imposing that each row in $d_w$ must have the same width as the corresponding row in $d$.
Let $\mathcal{F}_j$ be the set of indices corresponding to words following and in the same line of $d\idx{j}$ that are not untouchable:
\[
    \mathcal{F}_j \triangleq \{l \mid l > j \wedge d\idx{l} \text{ in same line of } d\idx{j} \wedge \neg u\idx{l} \} \, .
\]
Moreover, let $\mathcal{P}_j$ be the set of indices corresponding to words preceding and in the same line of $d\idx{j}$:
\[
    \mathcal{P}_j \triangleq \{l \mid l < j \wedge d\idx{l} \text{ in same line of } d\idx{j}  \} \, .
\]
Then, the total number of letters before $d\idx{j}$ (or $w$) is $P_j \triangleq \sum_{l \in \mathcal{P}_j} |d\idx{l}|$ (respectively $P_{wj} \triangleq \sum_{l \in \mathcal{P}_j} |d_w\idx{l}|$).

We could weight the to-be-maximised function with a coefficient $q$ defined as
\[
    \func*{q}{j} \triangleq \begin{cases}
    	0 & \text{if } |\mathcal{F}_j| = 0 \mathrel{\wedge} \\ & \phantom{if } \neg \left( P_j = P_{wj} \wedge |d\idx{j}| = |w| \right) \\
    	1 - \left|\frac{P_j + |d\idx{j}| - P_{wj} - |w|}{|\mathcal{F}_j|}\right| & \text{if } |\mathcal{F}_j| \neq 0 \\
    	1 & \text{if } |\mathcal{F}_j| = 0 \wedge P_j = P_{wj} \mathrel{\wedge} \\ & \phantom{if } |d\idx{j}| = |w|
    \end{cases}
\]
and impose that $q$ must be strictly positive, at least for the last word.
If the last word has no homograph neighbour that meets the requirement, we can choose a non-homograph one and renounce to embed a bit.
Even though such a stratagem undoubtedly increases the embedding effectiveness, there is still room for improvement.
A simplification that still yields acceptable results is to impose that the watermarked line must have the same length or be shorter than the original one.
Then, it is up to the \pdf{} embedder to produce a document with the required structure.

\begin{figure}[H]
	\noindent\Large{\texttt{Sphinx of black quartz, judge my vow.}}\hfill{}
	
	\noindent\Large{Sphinx of black quartz, judge my vow.}\hfill{}

	\caption{Pangrams of a mono-spaced (above) and a variable-width font. Characters such as `i' and `l' are narrower in the lower writing.}
	\label{fig:pangrams}
\end{figure}

The proposed approach works best with mono-spaced fonts, whose characters each occupy the same amount of horizontal space.
However, proportional fonts are more commonly used.
Thus, we should expect characters to have different sizes.
Consequently, the only reliable method to impose proper formatting is to choose the line length and the minimum and maximum inter-word spaces, parse the font to retrieve the character widths and calculate if the watermarked text is acceptable.
If it is not, then replace one or more words with ones that similarity-wise are slightly worse but fit better in the line.
This approach is computationally expensive, but it is the one with the highest embedding rate.
To achieve optimal results, kerning must be taken into account.
Otherwise, lines could erroneously be identified as unacceptable.
Still, discrepancies should be so small to have no practical consequences in the majority of documents.

\subsubsection{Extraction}
Let $\Call{neighbours}{x}$ be the function that returns all neighbours of $x$, let $\Call{homograph}{x}$ be the function that checks whether $x$ is a homograph and let $\Call{extractBit}{x}$ be the function that extracts the payload bit from a homograph $x$.
Let \Call{checkInserted}{} and \Call{checkDeleted}{} be the functions that check that at most $\lambda$ words were respectively inserted or deleted by the adversary\footnote{Obviously, we could also choose two different coefficients for insertion and deletion checks.}.
The latter also returns the number of deleted payload-bearing words, so that we can guess the erased payload bits.
Each guessed bit is correct with probability ½.

\begin{algorithm}[h]
    \caption{Linguistic extraction. Part 1: untouched words.}
	\label{alg:linguistic-extract}
	
	\DontPrintSemicolon{}
	\KwData{original document $d$ as an array of words;\newline
	    watermarked document $d_w$ as an array of words;\newline
	    array $u$ containing whether the corresponding word in $d$ can be replaced;
	    max number $\lambda$ of words that could have been inserted or deleted by the adversary;
	}
	\KwResult{extracted payload $p''$ (see sec. \ref{sec:ecc})}
	
	\BlankLine{}
	$p'' \gets [\ ]$\;
	$i \gets 1$ \tcp*{counter for $p''$}
	$j \gets 1$ \tcp*{counter for $d$}
	$l \gets 1$ \tcp*{counter for $d_w$}
	
	\BlankLine{}
	\uWhile{$l \leq |d_w|$}{
		\If{$u\idx{j} \vee d\idx{j} \notin G_w$}{
    		\uIf{$d\idx{j} = d_w\idx{l} \vee \phantom{a}$ \newline
    		        $d\idx{j} \in \Call{neighbours}{d_w\idx{l}} \vee \phantom{a}$ \newline
    		        $\exists r : d\idx{j}, d_w\idx{l} \in \Call{neighbours}{r}$}{
    			\pe{j}; \pe{l}\;
    		}\uElseIf{$x \gets \Call{checkInserted}{d\idx{j},\, d_w,\, l,\, \lambda}$}{
        		\tcp{untouchable word is skipped}
    			\pe{j}; \pe{l}[x+1]
    		}\uElseIf{$x,y \gets \Call{checkDeleted}{d_w\idx{l},\, d,\, j,\, \lambda}$}{
    			$p''\idx{i \ \KwTo \ i + y - 1} \gets \Call{random}{[0, 1]}$\;
    			\pe{i}[y]; \pe{j}[x]\;
    		}\Else{
    			\Throw exception \tcp*{Lost sync}
    		}
    		\Continue
    	}
    }
\end{algorithm}
\begin{algorithm}
	\caption{Linguistic extraction. Part 2: payload-bearing words.}
	\label{alg:linguistic-extract-2}
    
    \endWhile{}{
    	\uIf{$d\idx{j} = d_w\idx{l}$}{
    	    \tcp{adversary removed payload bit}
    		$p''\idx{i} \gets \Call{random}{[0, 1]}$\;
    		\pe{i}; \pe{j}; \pe{l}
		}\uElseIf{$\Call{homograph}{d_w\idx{l}} \mathbin{\wedge} \newline d_w\idx{l} \in \Call{neighbours}{d\idx{j}}$}{
			$p''\idx{i} \gets \Call{extractBit}{d_w\idx{l}}$\;
			\pe{i}; \pe{j}; \pe{l}
		}\uElseIf{$\exists r : \Call{homograph}{r} \mathbin{\wedge} \newline d\idx{j}, d_w\idx{l} \in \Call{neighbours}{r}$}{
		    \tcp{may be wrong if multiple $w$}
			$p''\idx{i} \gets \Call{extractBit}{r}$\;
			\pe{i}; \pe{j}; \pe{l}
		}\uElseIf{$x \gets \Call{checkInserted}{d\idx{j},\, d_w,\, l,\, \lambda}$}{
		    \pe{l}[x]
		}\uElseIf{$x, y \gets \Call{checkDeleted}{d_w\idx{l},\, d,\, j,\, \lambda}$}{
			$p''\idx{i \ \KwTo \ i + y - 1} \gets \Call{random}{[0, 1]}$\;
			\pe{i}[y]; \pe{j}[x]
		}\Else{
			\Throw exception
    	}
    }

	\BlankLine{}
	\KwRet{$p''$}
\end{algorithm}

Algorithm \ref{alg:linguistic-extract} (and its continuation, alg. \ref{alg:linguistic-extract-2}) extracts a payload $p''$ from the document $d_w$.
Error correction is left to external components.
The algorithm automatically recovers synchronisation if the adversary changes, inserts or deletes words.
When possible, it also tries to recover the removed payload.
Since we assume that adversaries want to preserve the original meaning, we check for word insertion before deletion.
We presume that a word was inserted if
\begin{gather*}
	d\idx{j} = d_w\idx{l+1} \vee \phantom{a} \\
	d\idx{j} \in \Call{neighbours}{d_w\idx{l+1}} \vee \phantom{a} \\
	\exists r : d\idx{j}, d_w\idx{l+1} \in \Call{neighbours}{r} \, .
\end{gather*}
The word-deletion criterion is equivalent.
Both criteria can be easily generalised up to $\lambda$ consecutive inserted or deleted words.
However, the algorithm loses synchronisation when the adversary consecutively inserts and deletes words.
It is possible to handle even those extreme cases; however, it may be advisable to manually check the document.
Using \nlp{} decreases de-synchronisation rate, but may require human supervision.

If synchronisation is lost with respect to $\lambda$, the algorithm throws an exception.
Then, the user can choose to either increase the parameter or manually edit the document.
We can also design a simplified version without \Call{checkDeleted}{} and \Call{checkInserted}{}.
Whenever synchronisation is lost, the algorithm assumes that a single word was inserted.
Even though this version is less accurate, it is easier to implement, as we do not need to construct insertion and deletion check.

If we want to use the embedding algorithm proposed for variable-width fonts, it becomes quite difficult for the extraction algorithm to determine whether a word bears a payload.
Wet paper codes solve this exact problem, as they are designed for extraction algorithms that do not know which elements carry the payload \cite{10.1109/TSP.2005.855393}.
However, wet paper codes are not suited to traitor tracing, as they assume a passive adversary.
Instead, we can use a synchronisation \ecc.

\subsubsection{Error Recovery}\label{sec:ecc}
The error recovery mechanism depends on the watermarking algorithm.
With the one that we initially proposed, it is sufficient to use a Reed-Solomon code with \bch{} view.
We define the extracted payload $p''$ as $p'' \triangleq p' \mathbin\Vert \Call{ecc}{p'}$.
The only drawback is that we must decide the number of correctable errors.
Since such figure is dependent upon the number of embeddable symbols it can be easily estimated.
We start with the number remaining touchable words that can be found in $G_w$ as \ecc{} size.
Then, we decrease it until the embedding is successful.

Unfortunately, the situation is more complex if we decide to embed justified documents that use proportional fonts.
Borrowing steganalisys's terminology, we can say that the transmission channel is unknown to the receiver.
However, since we assume an active adversary, we cannot use wet paper codes.
Instead, we have to use watermark codes, which are designed to correct substitution, insertion, and deletion errors at the cost of a sizeable overhead.
In recent years, several efficient implementations have been proposed.
In particular, the concatenated synchronisation error correcting code designed by \cite{10.1109/ICCC47050.2019.9064318} is quite promising.

\subsubsection{Tagged Documents}
We assumed that documents are not tagged, meaning that the algorithm does not know the specific sense of the words occurring in the document.
Tagging must be performed manually by the author (or a qualified user) and is excessively time-consuming to be done on every document.
However, authors may be interested in tagging specific documents or words in order to ensure that meanings are not altered.
Thus, it may be interesting to expand the watermarking algorithm in order to include tagged documents.

The graph is slightly more complex.
Starting from $G$, for each word we add a generic vertex representing the entirety of its senses.
Let $w$ be a valid word, then we create an edge linking $\langle w', s\rangle$ and $w$ for each vertex $\langle w', s\rangle$ representing a pair in the same synset of one of the meanings of $w$.
Each vertex is weighted according to the function
\[
    \Call{weight}{\langle x, s_x\rangle, y} \triangleq \frac{\sum_{s_y \in \func*{S}{y}} \Call{sim}{\langle x, s_x\rangle, \langle y, s_y \rangle}}{|\func*{S}{y}|}
\] 
where everything is defined as before.
Finally, we can delete the original edges.
The remaining portion of the graph creation process, the embedding algorithm and the extraction one are left untouched.

Watermarked documents produced from tagged data are usually of higher quality than those produced from non-tagged documents.
As mentioned, we cannot expect organisations to manually tag their sensitive documents.
However, several sense-identifying neural networks have been presented in recent years.
In particular, \ewiser{} was the first to reach the threshold of 80\% correctness \cite{10.18653/v1/2020.acl-main.255}, which is comparable to how humans perform.
Determining whether \ewiser{} and analogous networks actually improve the watermarking algorithm require further study.

\subsection{Structural}\label{sec:structural}
In the previous section we assumed that the original document is retained, thus we could use a non-blind watermarking algorithm.
Unfortunately, we cannot replicate the reasoning in this context.
Since the document is modified by the synonym substitution, we are forced to choose a blind algorithm.

The technique proposed by \cite{10.1109/ICDAR.2003.1227767} is an ideal candidate.
In \cite{10.1109/ICDAR.2003.1227767}, words are divided into classes and labelled based on width comparison with neighbouring words.
Then, lines are divided into segments of $s$ consecutive words such that the first and the last word of each segment are shared with its neighbours.
Segments are divided into classes and labelled based on the labels of the words they contain.
The same amount of information is carried independently by each segment in its inter-word space statistics.
The payload is embedded by shifting non-shared words in each segment.
Designing embedding and extraction rules must consider the $s$, the number of segment classes and the payload size.

Conversely to the linguistic algorithm, there is the concrete possibility that we can extract the payload from a single section.
According to the adversary model, if the adversary does not publish the entire documents, they cannot alter its visual appearance.
Therefore, we can design word classes based not only on relations with adjacent words, but also on relations with corresponding words in the original documents (e.g. if the word carries a payload bit).
Obviously, we cannot increase the number of classes ad infinitum, lest the decrease of embedding effectiveness caused by the under-representation of specific segment classes.

\subsection{Font-Based Algorithm}\label{sec:font-based}
There are several \pdf{} metadata-watermarking algorithms.
Some use comments, others use invisible objects, still others modify the embedded fonts.
A technique similar to \cite{10.1145/1161366.1161387} is ideal.
Not only it is resistant to printing, it also splits the payload among several characters.
Thus, it is not removed when the adversary extrapolates a portion of the document.
Conversely, techniques that change glyphs in specific locations, such as the first proposed in \cite{US20040001606A1}, are not well-suited to the proposed adversary model.
They do not withstand the extrapolation of a document portion and the synonym substitution is more robust when the adversary publishes the whole document.

If altering character shapes is not feasible, a non-printing-resistant alternative is changing the font's character map.
The payload can be represented as an integer, thus we can construct an equivalent self-inverting permutation as described in \cite{10.1145/1839379.1839402}.
Then, we can embed the payload by changing the glyphs' (or the character codes') order according to the permutation.
The technique is not as conspicuous as it may seem.
The \pdf{} standard specifies copy protection.
However, it must be enforced by a compliant reader.
Thus, a common stratagem is to alter the \texttt{cmap} so that relations between glyphs and character codes are scrambled.
Obviously, the text needs to be modified so that the document retains its original visual appearance.
The adversary cannot seamlessly swap the font with a similar-looking alternative, otherwise the text would be garbled.
Thus, they may settle for publishing an image of the document, preserving the structural watermark.

\subsection{Payload}\label{sec:payload}
Using the user \id{} as payload is the easiest way to perform traitor tracing with watermarks.
However, it has a serious drawback that cannot be overlooked.
Suppose that an attacker gains access to a user account.
Then, the real user can divulge any document, blaming the attacker for the leak.
Secure logs are a deterrent, but the user may be able to mount a convincing defence based on the severity of the attack.

The obvious solution is to change user identifiers after every attack.
In order to have certain accountability, all identifiers must be changed even if the intrusion was not successful.
However, it is a non-trivial logistical problem when the number of accounts is vast.
In some cases, the user \id{} is linked to external factors (such as the social security number), thus changing it is impossible.

Therefore, payloads containing only the user \id{} do not offer sufficient accountability.
At least the download timestamp must be included.
In the following sections two possible payload for traitor tracing will be presented, each with its own pros and cons.

\subsubsection{Log-Independent Payload}
In this section we will present a payload type that does not rely on external logging mechanisms.
As showed, we need to embed in a document at least the user \id{} and the time of download.
In order to increase embedding effectiveness, the payload must be as small as possible.
The user \id{} typically is decided by external or logistic requirements, thus we will focus only on the timestamp.
We can choose from several time encodings, but we will take into consideration only the two that require the least storing space.

\iso{} 8601 is the international standard that defines the textual representation of dates, times, and intervals.
There are several string formats, of which the most compact is \texttt{[YYYY][DDD]T[hh][mm][ss]Z}.
Storing it as an \textsc{ascii} string requires 105\bit, so using it in the payload is not feasible.
However, since we fix the format and impose that all times are \utc, we can omit the time \texttt{T} and time zone \texttt{Z} indicators.
All remaining characters are numeric, therefore we can encode each component separately as an integer.
In total, this technique requires 38\bit.
If we use only the last two digits of the year (\texttt{[YY][DDD][hh][mm][ss]}), we can decrease the space requirement to 33\bit.

Unix-like systems use Unix time, which is the number of seconds that have elapsed since \isotime{1970-01-01}{00:00:00}, leap seconds excluded.
Traditional Unix time encoding uses a 32\bit{} signed integer and can represent times from \isotime{1901-12-13}{20:45:54} to \isotime{2036-01-19}{03:14:08}.
It is a compact representation, but the maximum date is too close to be used in new applications.
Modern Unix systems use 64\bit, which can represent a much wider range of times.
However, it requires too much space to be used in the payload.
Since we do not care about past dates, the best solution is to use a 32\bit{} unsigned integer, which can store a time up to \isotime{2106-02-07}{06:28:15}.

Both solutions are non-standard to some extend, even though there are historical examples of Unix times stored in unsigned integers.
As such, the latter technique is the preferred one.
The former's only advantage is that it correctly identifies leap seconds, but in the majority of applications such a high sensitivity is not required.
We can use even less storage size by choosing an arbitrary epoch zero.
However, this approach requires external information, partly defeating the purpose of having a log-independent payload.

Since the user \id{} is fixed and the timestamp is known, an aggressor can perform chosen-message attacks.
It should not be a problem if the algorithm used to encrypt the payload is sound.
Nonetheless, we can add an 8\bit{} random integer to prevent the aforementioned attacks.

\subsubsection{Log-Dependent Payload}
Contrary to the log-independent one, the payload scheme presented in this section relies upon a log server.
We suppose that download logs are stored in a relational database that includes a table with at least the following columns:
\begin{itemize}
    \item \downid{} (\texttt{bigint}): \id{} of the download.
    It constitutes the actual watermark payload.
    It must be randomly generated, so that an aggressor cannot perform chosen-message attacks.
    \item \docid{} (\texttt{bigint}): \id{} of the document.
    It must uniquely identify a document, independently from the specific version.
    \item \texttt{user\_id} (\texttt{bigint} or \texttt{string}): \id{} of the user that downloaded the document.
    \item \texttt{timestamp}: date and time of download.
    Second precision is sufficient.
    \item \texttt{ip\_addr} (\texttt{inet} or equivalent): \textsc{ip}v4/\textsc{ip}v6 address from which the document was downloaded from.
\end{itemize}
This information allows to uniquely identify a download and provides all elements required to trace it back to the offending user.

Assuming that identifier and time fields are 8\byte{} long and that \texttt{inet} fields are 7\byte{} long for  \textsc{ip}v4 addresses (or 19\byte{} for  \textsc{ip}v6), then the total row size is 39\byte{} (or 51\byte).
Storage is relatively inexpensive, thus we can freely add other pieces of information that may be relevant in a later investigation (e.g. the user's security clearance).
The embedded payload contains only \downid.

When designing the log database, we can choose between two unique constraints.
The easiest and most reliable method is to choose \downid{} as (primary or unique secondary) key.
However, it greatly reduces the number of downloads we can watermark with a given \downid{} size.
Choosing a larger payload affects negatively imperceptibility and embedding effectiveness, which are the strong points of log-dependent payloads.
The other technique consists in choosing the combination \downid{}-\texttt{document\_id} as unique key.
It is more space efficient, but it increases the complexity of the watermarking extraction algorithm.
 
Log-dependent payloads are smaller than log-independent ones.
This characteristics plays a pivotal role on imperceptibility and embedding effectiveness in short documents.
As such, in some cases they are the only suitable solution.
Nonetheless, they rely on external servers.
If logs can be altered, high-privileged users can cover their tracks without performing illegal operations.
The problem is mitigated by a clear separation between users and system administrators.
However, it is not resolved: a user and an administrator can ally in order to circumvent the protections.
The only reliable solution is to make the logs unchangeable, so that all operations can be audited.

\subsubsection{Payload Choice}
We assume that the leaker wants to be credible.
They can either publish the entire document (after having performed formatting- and content-based attacks) or just an untouched section.

Let us analyse the former case.
If the document is subject to formatting-based attacks, we can assume that structural and font-based payloads have been removed.
Thus, it is indifferent which payload we choose for those two algorithms.
Since the burden of traitor tracing is placed entirely on the linguistic algorithm, it is advisable to choose the log-independent payload for it if the length of the document allows it.

Now, let us focus on the latter case.
Only a section of the document is published by the adversary, therefore the linguistic payload has been removed.
Given the structural algorithm's construction, a small payload is advisable.
Therefore, the log-dependent one is the best choice.
Conversely, the metadata-based algorithm does not have specific size constraints.
Thus, we are free to choose the log-independent payload in order to increase, although marginally, robustness.

\subsubsection{Encryption}\label{sec:encryption}
The watermarking algorithm must guarantee payload confidentiality and authenticity.
Therefore, we must adopt authenticated encryption (\aenc).
There are three main \aenc{} approaches: encrypt-then-\mac, encrypt-and-\mac{} and \mac{}-then-encrypt.
Since the first is the most secure \cite{10.1007/s00145-008-9026-x}, it is the one we will adopt.

The payload is encrypted using the function $\Enc{} : \mathcal{P} \times \mathcal{K} \to \mathcal{C},\ (p, k) \mapsto c$, then the message authentication code (\mac) is generated from the ciphertext by the function $\Tag{} : \mathcal{C} \times \mathcal{K} \to \Phi,\ (c, k') \mapsto \phi$.
The final payload $p' \triangleq c \mathbin\Vert \phi$ is just the concatenation of the ciphertext and the \mac{}.
In order for the authenticated encryption to be resistant to all kind of attacks, \Enc{} must be \indcpa\footnote{A cryptosystem is indistinguishable under chosen-plaintext attack (\indcpa) if every probabilistic polynomial time adversary has only a negligible advantage over random guessing.} and \Tag{} must be strongly unforgeable.

Authenticated encryption applies to both symmetric and asymmetric cryptography.
Both have readily available secure implementations.
Symmetric encryption is usually faster than asymmetric, although the latter may be preferable from an ethical standpoint (see sec. \ref{sec:ethic}).

\subsection{Chilling Effect}\label{sec:ethic}
A whistle-blower is a person that informs someone in authority that offences are being committed, especially in a government department or a large company.
In order for the whistle-blower to be credible, they must have compelling evidence to support their allegations, which the regulating body can use to confirm such claims.
A case would never continue on legally, or ever be reported by news services, without substantial documentation.

Even though whistle-blowers' identities are usually secret and persecuting whistle-blowers is illegal in many countries, being one is not risk free.
There have been several reports of whistle-blowers being retaliated against from those they accused or alleged of wrongdoing.
As a result, potentials whistle-blowers are often discouraged from reporting their concerns for fear of reprisals \cite{eu-2019-1937}.
The watermarking algorithm we present may become yet another impediment to whistle-blowing.

This is most concerning when the whistle-blower is employed in a government agency.
Those who may be adversely affected by allegations are typically those in a position of authority.
They are the same people that have access to the case files, therefore they can discover the whistle-blower's identity by extracting the watermark payloads.
The natural solution is to protect the payloads through asymmetric encryption.
The decrypting key should be known only by an external trusted entity, which must use it exclusively when ordered by an impartial third party (ideally a judge).
However, since there is currently no law that protects whistle-blowing identities from legitimate discovery attempts, the adoption of protection mechanisms rests entirely on the willingness of the management body.

\section{Implementation}
\begin{figure}[h]
	\centering
	\fbox{\includegraphics[width=.475\textwidth]{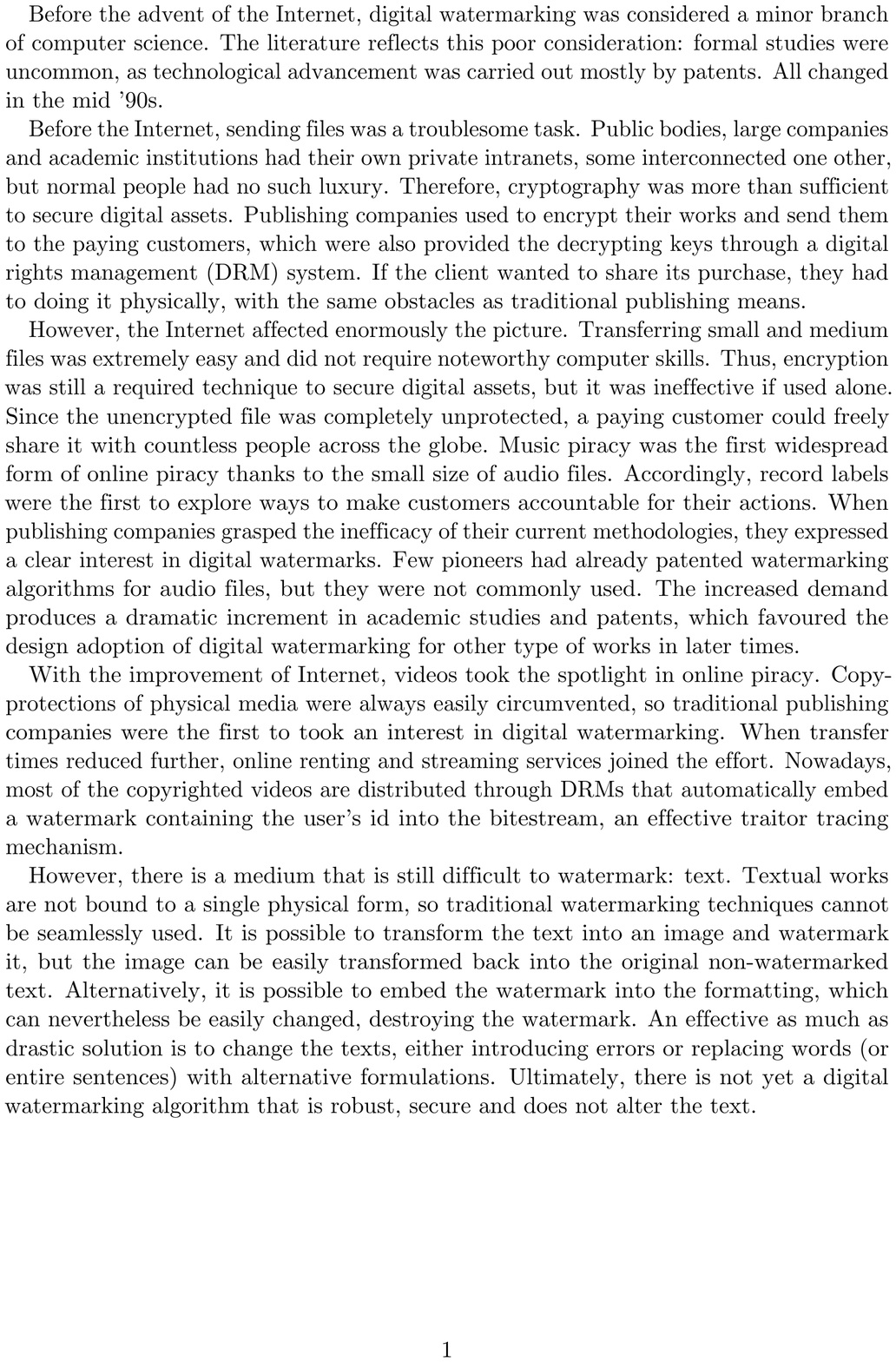}}
	\caption{A justified \pdf{} document.}
	\label{fig:test-document}
\end{figure}

\begin{figure}[h]
	\centering
	\fbox{\includegraphics[width=.475\textwidth]{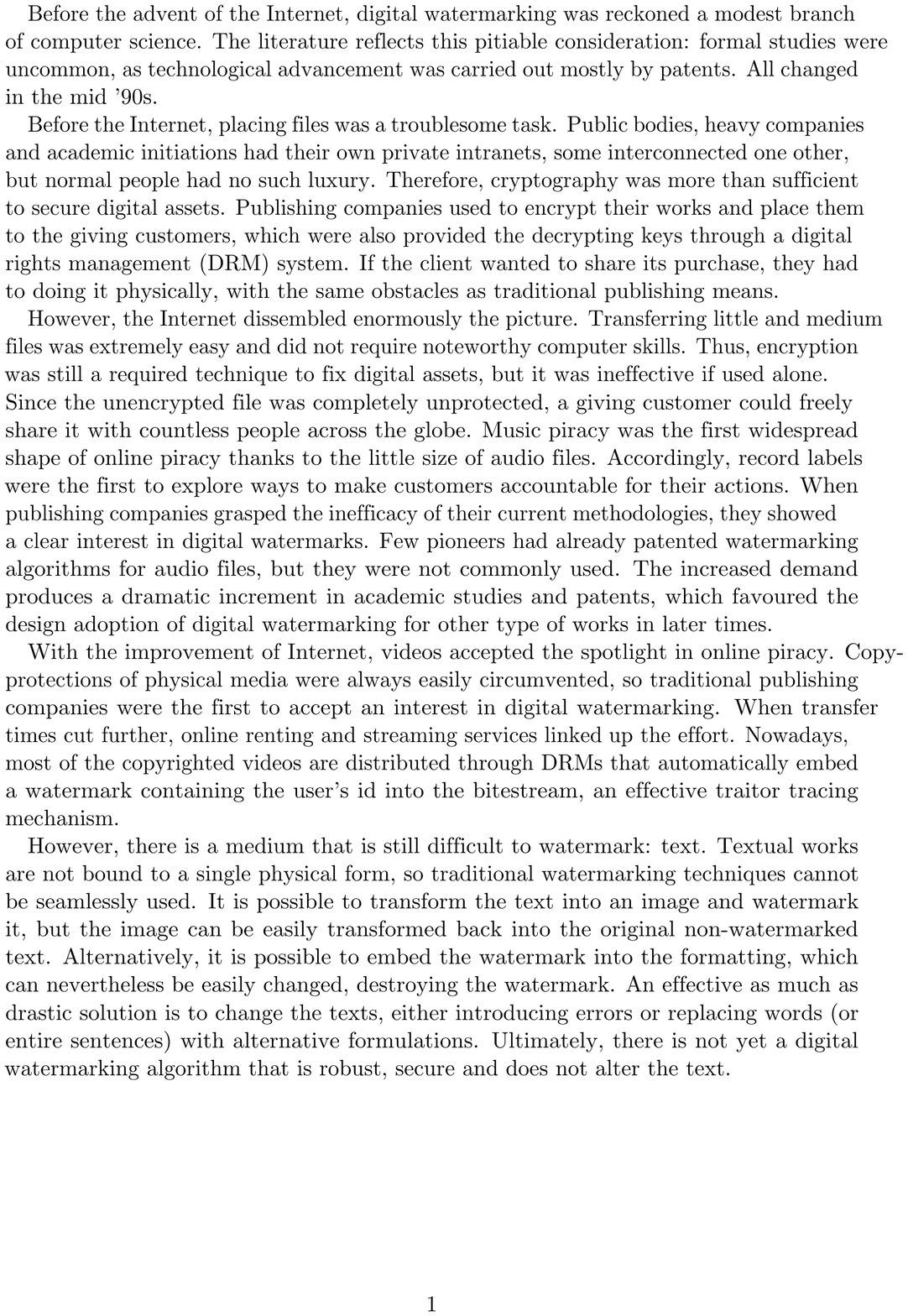}}
	\caption{A watermarked version of the document in fig. \ref{fig:test-document}. The payload is \texttt{0xA6A3CA}. Justification was not enforced.}
	\label{fig:test-document-marked}
\end{figure}

The implementation is composed by two parts: the watermarking algorithm in the strict sense and the \pdf{} parsing/embedding applicative.
The former is the same for all kinds of files, while the latter is dependent on the document formatting.
Figures \ref{fig:test-document} and \ref{fig:test-document-marked} show a \pdf{} document before and after the embedding of a 3\byte{} payload with no \ecc{}.

\subsection{Watermarking Algorithm}
We chose to implement the watermarking technique in Python thanks to its flexibility and its widespread adoption.
The \texttt{watermarking} package is composed by three distinct sub-packages, each corresponding to one of the presented algorithms.

Implementation of structural and \texttt{cmap}-based algorithms is straightforward and follows closely sec. \ref{sec:structural} and \ref{sec:font-based}.
On the contrary, implementations of the linguistic algorithm must include natural language processing.
My original intention was to use only spaCy \cite{spacy} with a WordNet plugin.
However, we had to adopt LemmInflect \cite{lemminflect} for lemmatisation and inflection and \nltk{} \cite{nltk} for word similarity.
We adopted \nltk{} rather than Wn \cite{wn} because the latter does not implement \ic-based similarity functions yet.
WordNet distinguished between several word classes, thus it is necessary to select only the neighbours with the correct part-of-speech (\pos) tag.
We can use \pos{} tags during payload extraction only if the adversary has the same knowledge of \nlp{} techniques.
Otherwise, we may wrongly identify a word as deleted.
An issue concerns the tokenisation of the document.
While phrasal verbs can be easily identified with \pos{} tags (\texttt{\textsc{rp}} in the Penn Treebank  \cite{penn-treebank}), generic multi-word tokens may be wrongly parsed as separate entities (e.g. the adverb `a lot' may be interpreted as the indefinite article `a' and the noun `lot').

In order to produce watermarked documents visually similar to the originals, to create an intermediate \pdf{} file after the execution of the linguistic algorithm is recommended.
In this way, the embedder can adjust the formatting before the second algorithm alters inter-word spaces.
We have the best results when the original document was automatically produces from a master template.
Then, the embedder can call the formatting applicative to produce a perfect document.

\subsubsection{Homograph Identification}
Homograph identification deserves a brief aside.
Trivially checking whether a word belongs to multiple synsets is formally correct but does not yield satisfactory results.
Let us analyse the verb `journey'.
Even though it belongs to two synsets, \texttt{travel.v.02} and \texttt{travel.v.04}, it is not a homograph.
Although the synsets have two different formal definitions\footnote{Respectively `undertake a journey or trip' and `travel upon or across'.}, they represent approximately the same concept and contain the same words\footnote{`journey' and `travel'.}.
Therefore, `journey' lacks the necessary ambiguity to be a payload-bearing word.
The culprit is WordNet's definition granularity, which is excessively fine in some context.
Inter-annotator agreement, the percentage of words tagged with the same sense by two or more human annotators, is estimated between 67\% and 80\% \cite{10.1145/1459352.1459355}.  

In our implementation, we consider a word as a homograph if it belongs to at least two disjoint synsets (i.e. two of its synsets have no words in common).
While this technique decreases false positives, it does not solve the problems of false negatives.
Let us consider the noun `street'.
It belongs to five synset, including the nearly identical `a thoroughfare (usually including sidewalks) that is lined with buildings' and `the part of a thoroughfare between the sidewalks; the part of the thoroughfare on which vehicles travel'.
However, none of the five sysnets contains words other than `street', not even `thoroughfare'.
This discrepancy is once again caused by WordNet's granularity.
Several coarse sense inventories have been proposed, but their use in the watermarking algorithms requires further study.
There is the risk of choosing an incorrect replacement if the granularity is too coarse or the inventory is not well-suited to the jargon used by the organisation.

\subsection{Parsing and Embedding Algorithm}
The parsing/embedding algorithm must adapt to document's typographical characteristics, e.g. section levels, paragraph modifiers and justification. 
Fortunately, all corporate \pdf{} documents are generated from a limited number of master templates.
A parsing and embedding algorithm based on a template's structure should work seamlessly with all derived files.
As a proof of concept, we decided to focus on documents composed by a single page with no elaborate formatting.
Nonetheless, the algorithm can be easily adapted to multi-page documents with a cycle on the pages.

\Pdf{} is a standard designed to display final products.
Thus, editing existing files is neither easy nor intuitive.
The libraries up to the task are quite scarce.
There are none in Python, the language in which we implemented the watermarking algorithm.
After careful consideration we opted for HexaPDF \cite{hexapdf}, a Ruby gem.

\begin{figure}[H]
    {\Large\texttt{[(\pdftoken{Before})\pdfspace{-312}(\pdftoken{the})\pdfspace{-311}(\pdftoken{advent}\pdfspace{-312}(\pdftoken{of})\pdfspace{-311}(\pdftoken{the})}}\hfill{}
    \Large{\texttt{\pdfspace{-312}(\pdftoken{Internet,})\pdfspace{-317}(\pdftoken{digital})\pdfspace{-312}(\pdftoken{watermarking})}}\hfill{}
    \Large{\texttt{\pdfspace{-311}(\pdftoken{was})\pdfspace{-312}(\pdftoken{considered})\pdfspace{-311}(\pdftoken{a})\pdfspace{-312}(\pdftoken{minor})}}\hfill{}

    \Large{\texttt{\pdfspace{-312}(\pdftoken{b})\pdfspace{1}(\pdftoken{ranch})]TJ}}\hfill{}

    \caption{
        The first line of the document in fig. \ref{fig:test-document}.
        Word fragments are in teal and spaces in orange.
        The last `b' is detached from the rest of the word due to kerning.
    }\label{fig:first-line}
\end{figure}

Given the simplicity of the document, the parsing mechanism is not complex.
Each \texttt{TJ} operator is analysed separately, then results are coalesced into a single array.
Word fragments and inter-word spaces are extracted with regular expressions based on parentheses.
We chose an arbitrary boundary of 200\pt{} between kerning and inter-word spaces.
Misclassified instances are improbable: the former is two orders of magnitude shorter than the latter.

\section{Conclusions and Future Works}
Protecting sensitive data is a vital endeavour for companies and government agencies.
To this end, they adopt multiple security mechanisms to defend themselves from external and internal attacks.
However, employees are harder to control because they have legitimate access to the confidential documents they may divulge.
Therefore, any document-sharing service should include a traitor-tracing mechanism, mainly to act as a deterrent: preventing leaks is much more desirable than bringing the source to justice.

In this paper, we presented an invisible watermarking algorithm for \pdf{} files that can be used in traitor-tracing mechanisms.
The technique is robust with respect to adversaries that wants to be credible (e.g. newspapers).
However, unwanted consequence may arise following the adoption of the presented framework.
It may cause a chilling effect, which is the inhibition or discouragement of the legitimate exercise of natural and legal rights by the threat of legal sanction.

From a technical standpoint, there is margin for improvement.
The lexical database cannot be easily updated.
This is troublesome for expanding organisations, in which the specialised jargon may shift.
Moreover, the natural-language-processing portion of the linguistic algorithm is not in par with the state of the art.
The most important issue is that components of multi-word elements may be wrongly recognised as separate (e.g. `a lot' is parsed as `a', `lot').
Fortunately, \nlp{} algorithms are quite apt at identifying phrasal verbs.
Another serious problem is that the chosen homograph is not always the most apt to the sentence.
Integrating a word-sense disambiguation technique such as \ewiser{} could be a major enhancement.
In addition, in order to mitigate WordNet's excessively fine granularity, new sense inventories specific for synonym substitution should be created.

\bibliographystyle{splncs04}
\bibliography{paper}
\end{document}
\endinput